# Universal Behaviour of Settling and spreading of particle clusters in quiescent fluids in confined vessels

by

Sayan Pal[a,b] and Amol A. Kulkarni[a,b]*

[a]Academy of Scientific and Innovative Research (AcSIR), CSIR-NCL Campus, Pune, India

[b]Chem. Eng. & Proc. Dev. Div., CSIR-National Chemical Laboratory, Pune - 411008, India

*Corresponding Author
Phone: +91-20-25902153, Fax: +91-20-25902621, E-mail: aa.kulkarni@ncl.res.in

**Abstract**

Here we report experiments on particle cluster settling at high Reynolds number in quiescent fluid contained in a vessel. The particles were observed to settle in a near-circular shape irrespective of the shape of the vessel cross-section and particle shape, size, and types. Effect of different parameters such as mass, type and aspect ratio of the particles, height, and viscosity of liquid was investigated. Formation of the hemispherical bottom cap of the cluster that bounces upon hitting the vessel bottom surface was found to be responsible for the final circular shape of the settled structure. Particle leakage from the cluster was seen in the form of a tail. In the liquid having viscosity beyond 100 cP, cluster breakage was observed that resulted in hindered settling and asymmetric shapes of finally settled particles. The observations are useful to understand the overall area over which settling of such clusters can be observed.

**Keywords:** particle cluster, temporal evolution, settling, spreading

# 1. Introduction

The motion of single particles or particle assemblies in a liquid in a constant gravity field has been studied extensively over the last 170 years.[1-5] Several processes involving the motion of particle clusters in liquids are found in nature. Such systems are pertinent to the dynamics of particulate gravity currents which have applications in oceanographic (turbidity currents and study of carbon fluxes to the sediments in the ocean, lakes and reservoirs), geological (debris flows, pyroclastic density currents; lava flows from volcanic eruptions), and environmental (pollutant-laden wastewater treatment) scenarios. Moreover, the settling study of such particle-fluid systems has significant commercial applications in fields as diverse as effluent dispersal, food processing, dredging waste removal, and mine tailings.[6-10] Understanding the nature of settling and spreading of such clusters in water is also important for undesired situations like airborne accidents of flights over the sea as rapid settling can restrict the area of investigation. This work performed using controlled experiments gives some insights in cluster sedimentation and settling in liquids.

When a swarm of particles is allowed to settle in quiescent fluids, it forms a cohesive entity called as cluster (also referred as blob or drop).[1, 11-14] The motion of the cluster is often considered to be analogous to the settling of a drop of a viscous fluid in a comparatively lighter liquid where a jump in the value of the particle concentration is related to the liquid-liquid interface.[1, 4, 12, 14] The ambient flow within such a settling cluster geometry was reported to be similar to the toroidal circulation observed in a settling liquid drop.[15] The behavior of a cluster of particles is different from a single particle in such scenarios, mostly because of the relative motion of particles in the cluster and the relative motion of clusters. The liquid motion at any point in the cluster is dependent on the relative velocity of the constituent particles and how distant the particles are from the point. Each constituent particle carried some amount of fluid along with it and convected by the velocity fields produced by the other constituent particles. This results in velocity enhancement of constituent particles and of the whole cluster as well.[1] The fall velocity of the cluster was observed to be higher at all times than the individual particles, and the enhancement in the falling rate increased when the particles were more closely packed[3]. Settling velocities of particle aggregates were found to be 4 to 8.3 times higher than the stokes law model. As the cluster takes a macroscopic identity, cluster Reynolds number ($Re_c$) based on the macroscopic scales, i.e., cluster radius and velocity, was often used for flow characterization instead of the particle Reynolds number.[1, 12, 14] Moreover, the added mass force plays an important role in the overall force balance.

At very low/zero $Re$, a cluster would maintain spherical shape while with increasing Re significant temporal evolution of the cluster is observed. For a certain time, depending on the specific system, the particles recirculate and stay together within a single almost spherical cluster/blob, after which the cluster forms a torus, then a ring-like structure and then eventually disintegrates. This phenomena has been reported by several authors in different settling systems over a range of Reynolds number (i.e. $1 \times 10^{-4}$ to 10) for particle sizes over a range of few microns to millimetres. [1, 4, 14] The particles were observed to move with respect to each other inside the cluster, and the cluster significantly changed its shape [12] while responding to the various forces acting on it. Machu et al. [14] reported the lower part of the cluster takes the shape of a roughly hemispherical cap, while the upper part resembles the conical shape of the laminar jet formed during the injection process. At comparatively higher Re, i.e., 93 to 425, Daniel et al. [6] observed that the variance of the cluster diameter grows quadratically at the beginning and attain a slower sublinear regime after some time. The temporal evolution of cluster was reported to be significantly dependent on the cluster $Re$ and the initial number of particles in the cluster.

During the past few decades, there have been several notable experimental as well as theoretical advances in the understanding of the dynamics of motion of cluster/blob made of particles made of micron to milli meter ranges. [1, 3, 12] These investigations were focused on $Re \leq 1.0$ in the creeping flow regime, where the inertia of particles and fluid and hydrodynamic interactions can be neglected. For $Re > 10$), both viscous, inertial, and buoyancy effects become significant and relative motion of the particles increases.[3] Wake mediated interactions starts becoming dominant with increasing $Re_c$ of the system. [6, 16] It can be said on the basis of the reported literature that beyond the creeping flow regime, the hydrodynamic interactions between particles mediated by the fluid become non-linear and demonstrate complex dynamics, for which our understanding of such systems is still rudimentary.

In the recent years, considerable attention has been focused on understanding the motion and evolution of fluid-particle systems such as a spherical settling cluster at low Reynolds numbers [1, 4, 6, 12-15, 17] but the spreading behavior of particle cluster upon hitting the bottom surface of the vessel apparently received little attention. Daniels et al. [6] only reported that even a localized release of particles into a quiescent fluid might give rise to sediment spreading on a greater section of the vessel bottom surface due to radial cluster expansion resulting from the source-flow interactions of the particles. To the best of our knowledge, to date, a detailed study of the corresponding observations of particle spreading from the clusters is not reported. Instead of settling at the centre of the vessel, particle clusters were observed to form a ring-like structure around the vessel centre in our experiments. Therefore the goal of this paper is to perform a detailed experimental analysis of settling at short time scales and spreading nature of the solid particle clusters in quiescent fluids in confined vessels. The focus is to investigate the solid deposition patterns at the vessel bottom surface. The effect of different parameters such as vessel geometries, wall interactions, and fluid properties over the settled geometry was studied experimentally. This is a purely experimental work, and in-depth theoretical understanding by means of numerical simulations is out of the scope of this study.

In view of this detailed introduction, the manuscript is organized as follows: the next section details out the experimental methods and the implemented data processing techniques in this work. Subsequently, a description of cluster settling and spreading phenomena and effects of different parameters on the spreading behaviour were discussed in detail. Finally, important findings are summarized.

2. Experimental

2.1. Experimental setup

Particles filled in a small cylindrical cavity were released instantaneously at the center of the vessel filled with a quiescent Newtonian liquid from a certain elevation and were allowed to settle. The particle trajectory and settling pattern at the vessel bottom surface were monitored. Glass vessels of different sizes and shapes were used for performing the experiments. An in-house developed motorized system was used for the release of particles to eliminate manual variations in particle release. Particles were loaded into a cylindrical reservoir, and a thin plastic sheet holding the particles was slid instantaneously using a servo-controlled motor to release all the particles so that the cluster descents in the vessel. Schematic of the experimental setup is shown in Figure 1. The particles travel about 5 cm in the air before entering the liquid. The time needed to release the particles was less than 50 ms. Due to the rapid movement of the sheet, no lateral movement of particles was observed, and this technique significantly reduced the entrainment of air bubbles inside the cluster. Particle trajectory inside the liquid phase after feeding during its settling and the final settling pattern formed was monitored using a high-resolution Sony camera. The evolution of the settling cluster was observed via an angled mirror placed directly below the cluster's path of travel. The mirror was kept at 45° angle at the bottom of the vessel to observe the side-view and bottom-view of the glass vessels. Visualization of settling and spreading time and pattern of the solid particles in water was performed by video processing (Virtual Dub and iMovie software), and image analysis was performed using Image J and Matlab softwares.

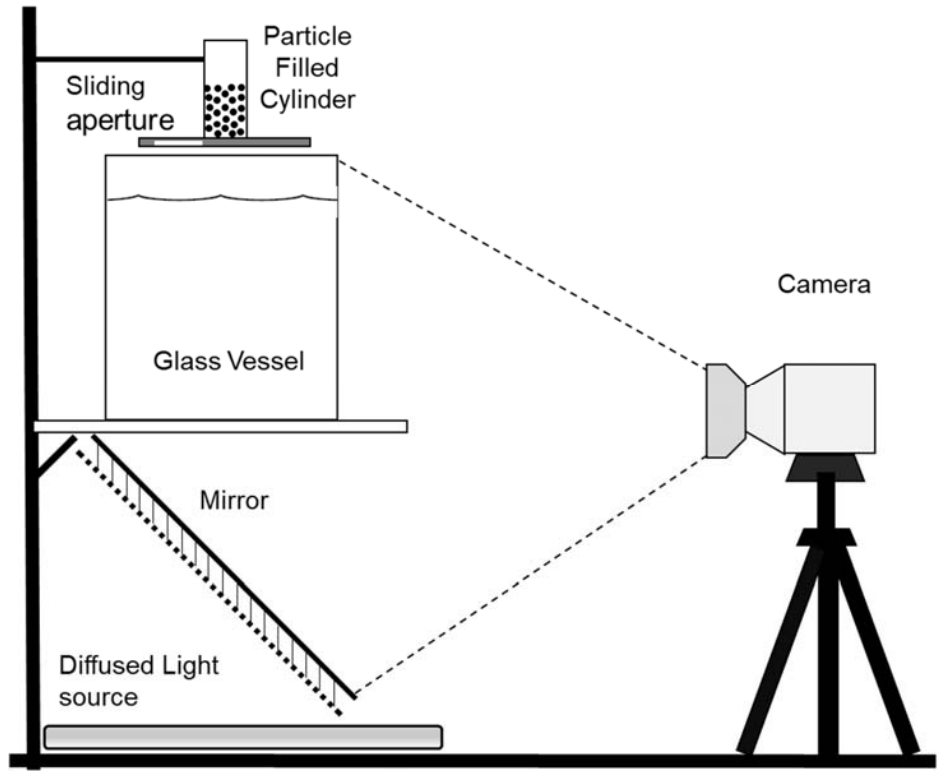

Fig. 1. Schematic illustration of the experimental setup used for settling of solid clusters. Particle settling and spreading was acquired as video in a mirror.

Most of the experiments were carried out in a 15 cm x15 cm x 20 cm square cross-section glass vessel using cubical sugar particles of 2.4 mm average size. The effect of different parameters was studied. These include: (i) initial mass of particles (6 gm to 16 gm), (ii) liquid height in the vessel (6 cm to 14 cm), (iii) particle size and type (sugar, glass and rice particles of different size and aspect ratio, see fig. 2), and (iv) viscosity of the liquid (1 cP to 700 cP). Vessels of different shape cross-sections viz. square, triangle, hexagon, cylinder in two different sizes were also used for performing experiments while other parameters were kept constant. Solutions of UCON (lubricant from Dow Chemicals) in water in different ratios were used for varying viscosity of the fluid without changing the density. The fluid viscosities were measured by a cone and plate viscometer.

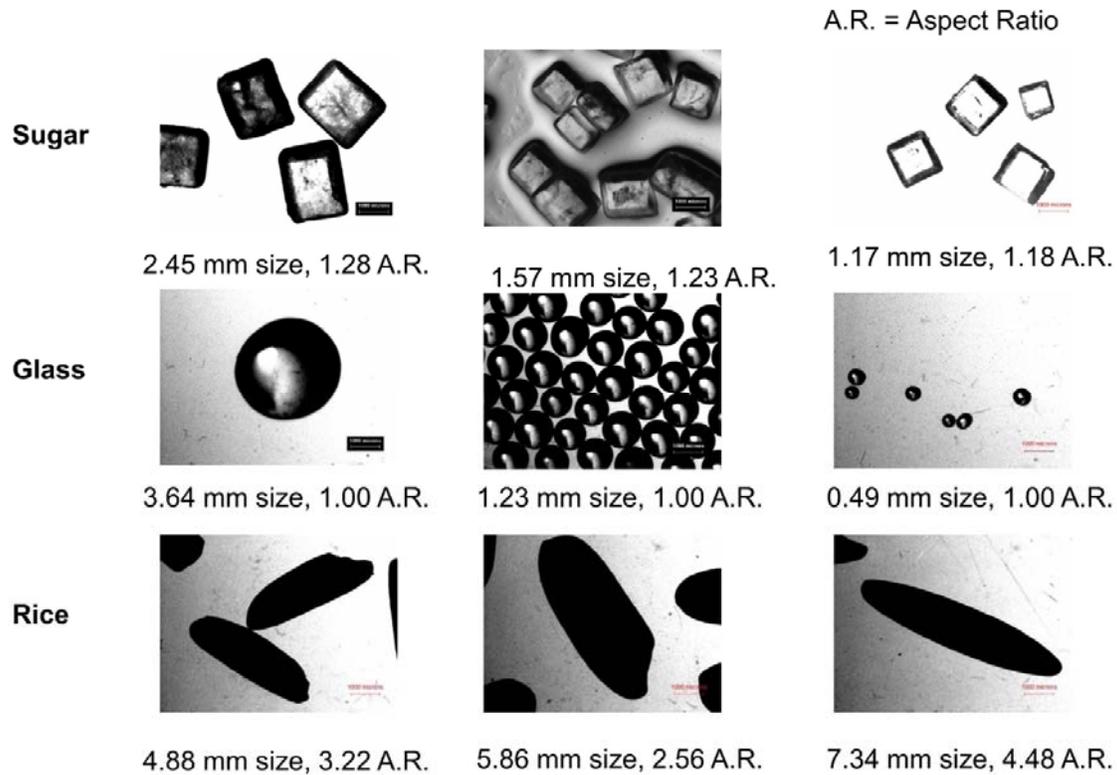

Fig. 2: Particle types and particle sizes used for settling experiments

## 2.2. Analysis

Settling time, spreading time, average spreading size (viz. diameters of the inner and outer circles, for the cases where it forms a circular shape) were measured as a response. Settling time was defined as the time needed for the cluster to reach the bottom of the vessel from the time it enters the liquid phase and spread time was defined as the time required to form the final settled shape after the cluster reaches the bottom surface. The time-averaged cluster settling velocity was determined as the total vertical displacement of the cluster divided by the settling time. The average diameter and roundness of the settled structure were determined by the minimum circumscribed circle (MCC), defined as the smallest circle, which encloses the whole of the roundness profile and the maximum inscribed circle (MIC), defined as the largest circle that can be inscribed inside the roundness profile. A schematic representation of the same has been shown in fig. 3. The average diameter and roundness were charecterised as the mean and ratio of the MCC and MIC, respectively. The diameters are presented in a dimensionless form throughout the manuscript by dividing it with the hydraulic diameter of the vessel cross-section.

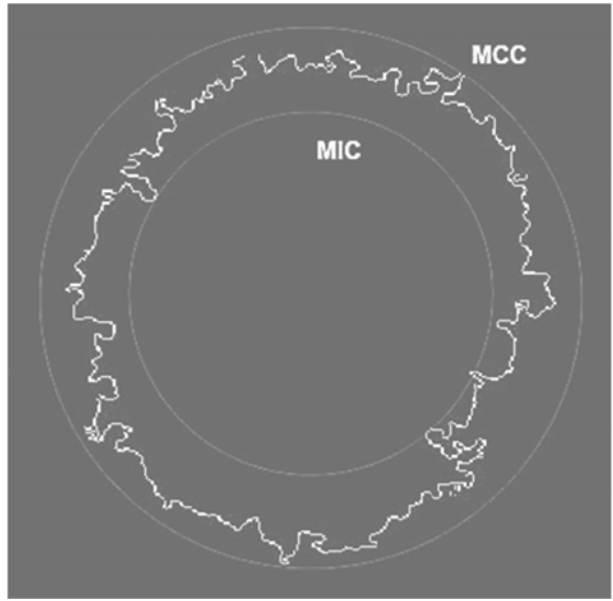

Fig. 3. Schematic representation of minimum circumscribed circle (MCC) and maximum inscribed circle (MIC)

As different types of particles having a wide range of aspect ratios were used for the experiments, the equivalent diameter of the particles [18] was used to characterize the particle length scale instead of particle size. The cluster Reynolds number ($Re_c$) was estimated as $(\rho_f V_{Cluster} D_{cluster})/\mu_f$.

## 3. Results and Discussion

### 3.1. Settling of clusters

While falling under gravity in a quiescent fluid, the group of particles is found to form a compact cluster under a wide range of conditions. The cluster of particles was seen to rapidly evolve into a nearly spherical shape. It is known from the literature that the motion and velocity of a settling particle cluster are different from the individual particles in the creeping flow regime. [1, 12, 14, 19] In our experiments, the settling velocity of the cluster never reaches terminal settling velocities of the individual particles (0.33 to 0.93 times the terminal settling velocity for different particle size, shape, and types) or the terminal settling velocities of the clusters (assumed spherical and treated as single particles) (0.28-0.5 times). In the experiments by Daniels et al.[6] performed outside the creeping flow regime, the settling velocity of the cluster and individual particles are reported to be a similar order of magnitude. However, in the present work, the cluster traveling time in the liquid was significantly lesser (settling distance is 1.71 to 5 times the cluster diameter compared to a minimum of 9.38 times in Daniels et al.[6]). This explains the significantly lower cluster velocities in comparison to the respective terminal settling velocities.

All the experimental data was carefully analysed using the images obtained from a high-speed camera (Kronos, Canada) at 1000 fps. to monitor the transient variation in the velocity of cluster during its settling. Since the impact of the cluster on the air-liquid interface followed by continuous displacement of water actually helped in generating a vortex with transient variation in the fluid velocity. Typical variation in the transient cluster velocity is shown in Fig. 4, which includes the settling velocity of the cluster from the time it touches the air-liquid interface, followed by its motion through the liquid and finally the collision with the bottom wall. Three different regimes are evident in the travel path of the cluster. Initially, the velocity continues to increase due to inertia until a hemispherical cap gets formed in the lower half of the cluster. Then velocity gradually decreases due to loss of energy while crossing the interface as well as due to viscous dissipation of cluster momentum. When the cluster while still in suspended form reaches close to the bottom surface some particles come apart from the cluster which momentarily increases the velocity of the cluster. Finally, a sudden drop in velocity is observed due to the buoyancy of the liquid coming from the bottom wall surface that leads to disintegration of cluster eventually subjecting individual particles to the motion. The time needed for the viscous dissipation region until the cluster was intact was the maximum among the three regimes. The time scale for the initial rise in the velocity was seen to be a function of the mass of cluster, while the time scale for viscous dissipation of cluster energy was seen to depend primarily on the liquid level. The time scale for the regime of cluster fragmentation was always the smallest among all. This particular trend was observed for all the experiments. The gradient of the velocity with time was seen to be a typical quadratic polynomial having two points of inflection and no symmetry.

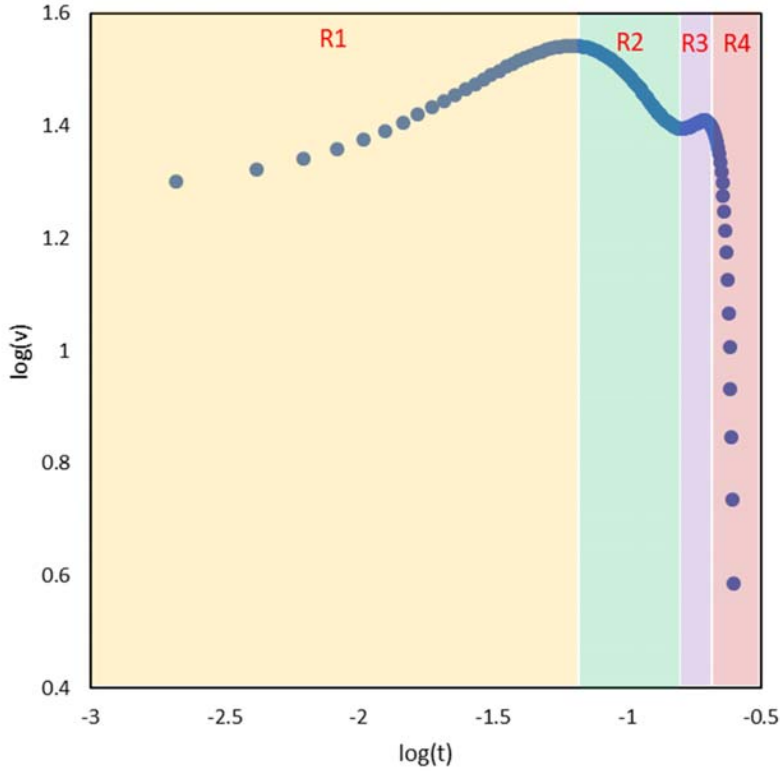

Fig. 4. Variation in cluster velocity as it travels through the liquid. Slope of different regimes - (R1- 0.142857, R2 -0.52632, R3 0.14, R4 -20)

The cluster diameter in our experiments was in the range of 2 cm to 4.6 cm. The recirculation of particles was observed in the cluster, which also provided a macroscopic identity to it. We have explored this feature by observing uniformly mixed distribution of settled particles of different colors when the reservoir was filled up in layers with different colored particles. While settling in the liquid, the lower part of the particle cluster forms a roughly hemispherical cap. While most of the particles were retained together, a few particles tend to leave the cluster during settling and remain in the tail portion, which is known as particle leakage (see fig. 5a). [1, 15] Particle leakage from the cluster primarily occurs because of the instabilities in the cluster when interacting particles undergo transient variation in the velocity, trajectories, and waves generated in the fluid due to entry of the cluster. The particles which are located in the outer layer of the toroidal circulation, tend to escape from the toroidal circulation and are dragged by the outside flow to form a tail of particles at the rear section of the cluster (see Fig. 5b). With the increasing initial mass of the particles, the tendency for particle leakage is observed to decrease and the clusters move downward with lesser deviation to its initial configuration. This is due to the lesser departure from closed streamlines of the toroidal circulation with increasing number of particles.[4] An important quantity which characterizes the near-spherical cluster formation is a critical mass (or number of particles) required to form a coherent spherical cluster. We had observed that when the mass of cluster was lower than the critical mass, the clusters were having a distorted/asymmetric shape and tended to disintegrate significantly and rather rapidly before reaching the vessel bottom.

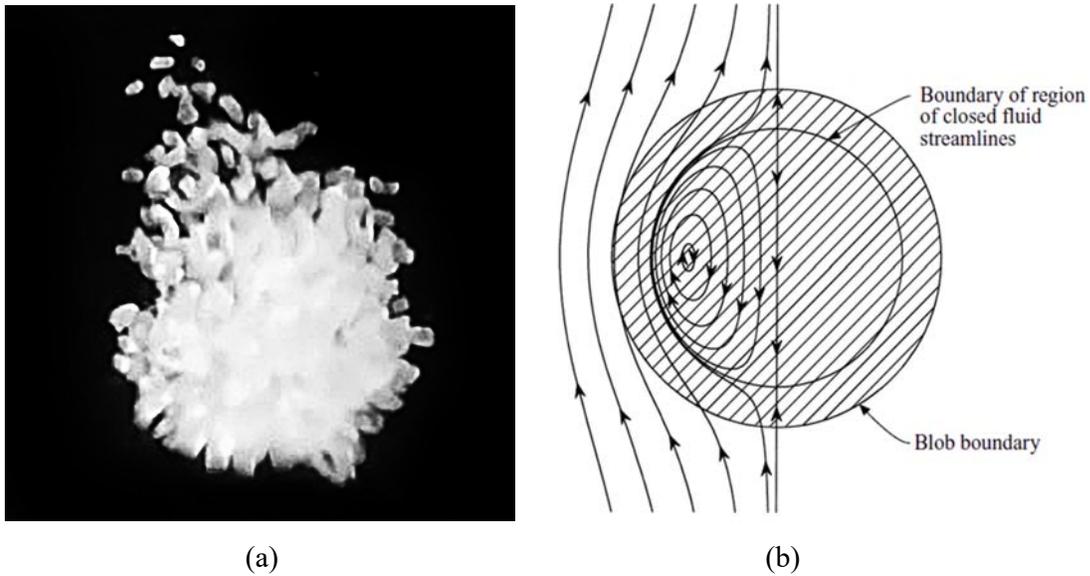

(a)  (b)

Fig. 5. (a) Settling cluster with tail formation at the back; (b) Schematic of open and closed-loop streamlines of toroidal circulation - reason behind Particle leakage and tail formation (Reprinted from Nitsche et al. 1997 [13] with permission from Cambridge University Press)

### 3.2 Spreading of clusters at vessel bottom

The entrance of the cluster in the liquid created ripples on the water surface, and as it travels towards the vessel bottom, the displacement of liquid in the downward direction along with the cluster helps generate a temporary circulation in the liquid. A strong interaction between sediment transport and the fluid dynamics due to the spatiotemporal displacement of water was observed during the spreading and settlement of the cluster particles. As the cluster hits the vessel bottom, the constituent solid particles spread out almost symmetrically under the influence of particle driven gravity currents under the balance of inertial and buoyancy forces. Particle driven gravity currents are generated by the release of a swarm of particles along with the interstitial liquid into a lighter ambient fluid. In such a scenario, the flow is driven by the difference between the bulk density of the particle current and the density of the ambient fluid in the vessel. The length of the particle gravity current is determined by the balance of buoyancy and inertial forces. The current length is very much higher than its thickness. Though it is reported that a propagating gravity current without particles shows a uniform velocity profile for $Re \gg 1$, when particles drive the flow in addition to the advective effects, they fall out of the flow losing some of the energy to viscous dissipation, and the buoyancy force continually decreases.[10] First, the particulate current propagates through an inertia dominant regime, then as the current reaches a certain length, the velocity and height at the tip of the current decreases. As a result, the viscous forces acting along the vessel bottom surface become more important, and the carried particles are tended to settle from the current there.[20] Fig. 6 shows the sequences of the cluster before it settles at the bottom wall surface. In the following section, effects of different parameters viz. initial cluster mass, vessel liquid level, liquid viscosity, vessel shape and size, particle size, shape, and type on the spreading behavior has been presented and discussed in detail.

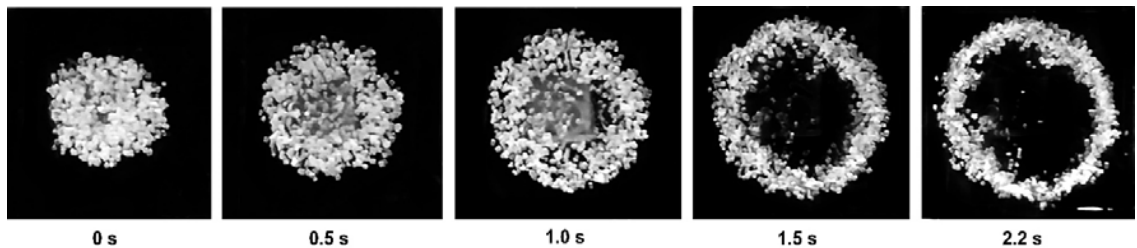

Fig. 6. Sequence of cluster spreading after collision with the vessel bottom wall. The zero time scale starts from the instant when the cluster touches the bottom of the vessel. A cluster comprising of 10 gm of 2.45 mm sized sugar particles was allowed to settle in a 15 cm X 15cm X 20cm square cross-section vessel filled with water (14 cm height).

### 3.2.1 Effect of the mass of particles

Due to the interparticle interaction in water and due to viscous effects, solid particles come near to each other and form a nearly spherical shape. A variation in the mass/number of solid particles affects the settling and spreading time. Increasing the mass of the solid particles resulted in a decrease in the settling time and an increase in the settling velocity due to higher inertia. Here the effects of added mass come become critical, which is the additional mass that an object appears to have when it is accelerated relative to a surrounding fluid.[21] With an increasing number of particles the added mass increases and adds to the effective inertia.

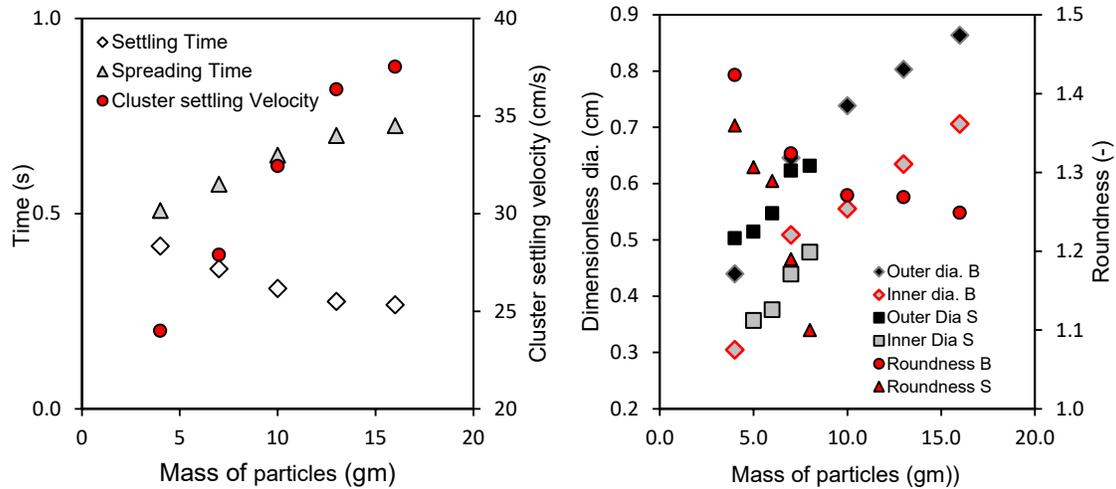

Fig. 7. Effect of the total initial mass of particles on (a) Settling and spread time in big square vessel; (b) Dimensionless diameter of the settled shape in small and big square shape vessels (B- big square vessel, S- Small square vessel)

The spreading time, as well as the average inner and outer diameter of the final shape of settled particles, were found to increase with increasing mass of solid particles irrespective of different vessel sizes (see fig. 7b). With an increasing number of particles and the size of the cluster, the bulk density of the resulting particulate current increases. This increases buoyancy, which is the driving force for the spreading of particles. With the increasing initial mass of particles, the roundness of the settled structure decreases, i.e., the settled shape becomes closer to a circle.

### 3.2.2 Effect of vessel liquid level

The settling and spreading time were observed to increase with the increasing liquid level. This is due to the increased hindrance provided by the liquid on the motion of particles. This hindrance is due to pressure force and also the viscous force. Though Bonnecaze et al. [20] found that lengths of the particulate currents in shallow surroundings were greater than compared to deep surroundings, Fig. 8 shows that the variation in liquid level has no significant effect on the inner and outer diameter of the settled shape. The roundness value of the settled shape tends closer to unity with an increasingly fluid level, which represents that the settled shape becomes more symmetric. Higher settling time provides more time to recirculate and forms a cluster of symmetric shapes. When the cluster shape is closer to a perfect sphere, with an impact on the bottom surface, the particles settle more symmetrically to form a near circle geometry.

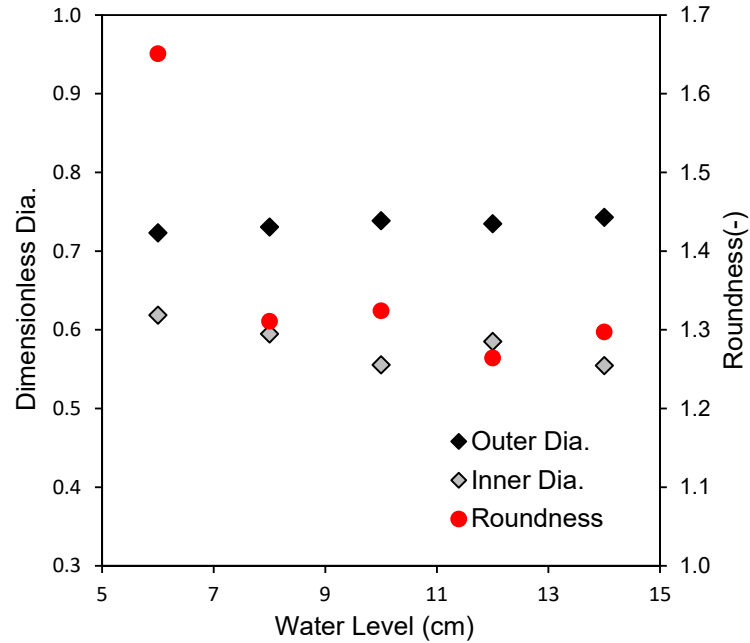

(a)

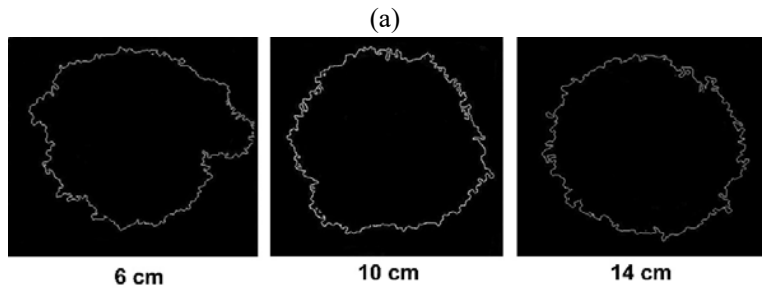

(b)

Fig. **8.** (a) Effect of the vessel fluid level on the settled pattern in the big square-shaped vessel, (b) Outer diameter of the settled shape for different vessel fluid levels.

### 3.2.3 Effect of Vessel shape and size

Vessels of different geometries in two different sizes were used to carry out particle settling experiments. It was interesting to observe that the settled shape is found to be close to a circle despite different shapes of the used vessels (see fig. 9). As the water level was kept constant in all the experiments, the liquid volume in the vessel was dependent on the size and shape of the bottom surface. As discussed in section 3.2.1, the threshold value for the mass of particles to form a symmetrical settled shape is dependent on the size of the vessel bottom surface area or the proximity of the vessel wall. Since it was observed that the dimensionless diameter increases linearly with increasing mass of particles (see section 3.2.1), in order to understand the effect of vessel shape and size, the outer diameter used in the following analysis was normalized by the mass of particles used for the experiments. Fig. 10a shows an interesting observation that irrespective of the vessel shape, the normalized diameter decreases with increasing vessel volume. It is also evident from the fig. 10b that the square shape vessel results in enhanced spreading of the particles in comparison to other vessel shapes.

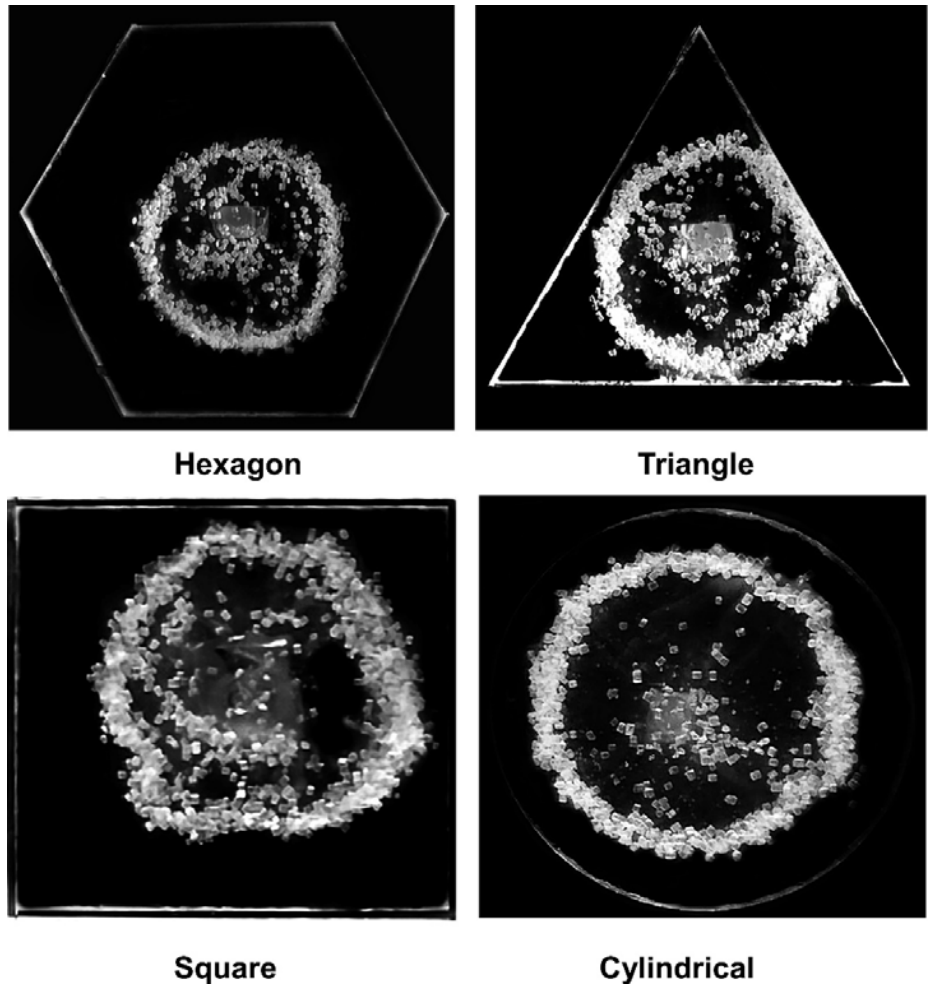

Fig. 9. Effect of vessel shape on the settled pattern

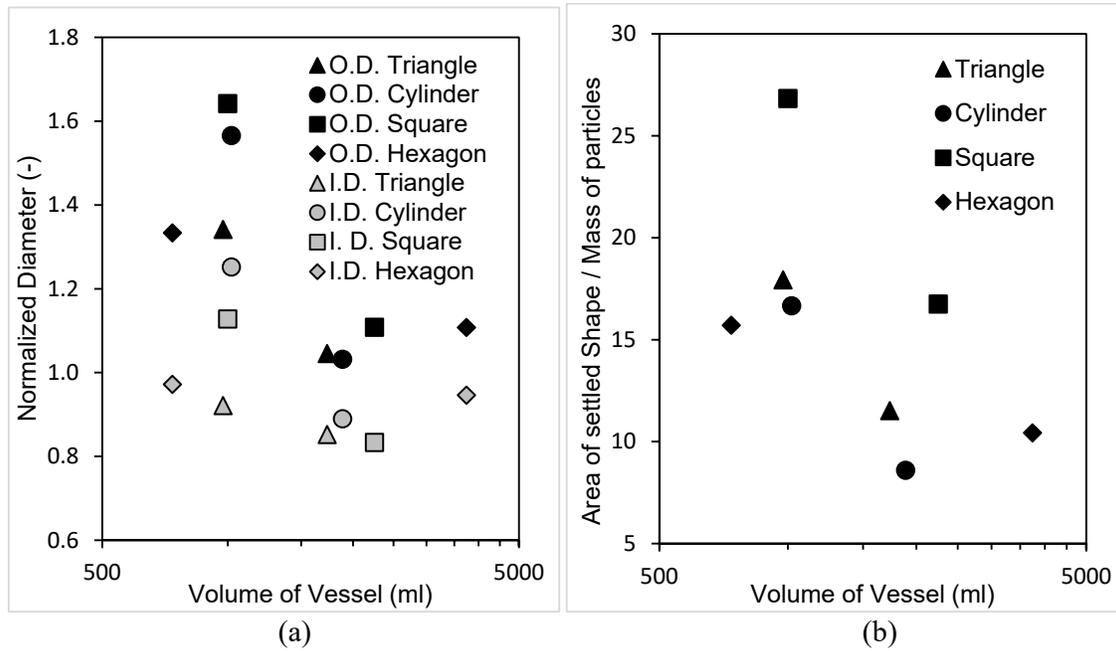

Fig. 10. Effect of vessel shape and size on the dimensionless diameters of the settled pattern.

### 3.2.4 Effect of particle size, shape, type

To understand the effect of particle type, size, and shape on the settling phenomena, different particle types of various sizes and aspect ratios were used for the settling experiments when other parameters were kept constant. Fig 11a shows no significant deviation was observed in the dimensionless outer diameter of the settled shape for particles of various diameters (ranging from 492 μm to 3643 μm average equivalent diameter). In general, at ambient conditions, the settling velocity of a particle decreases if the shape of a single particle deviates from a sphere. [22]. The drag force on the individual particles would change with a significant variation in the aspect ratio. However, despite these facts on the settling behavior of individual particles, it was interesting to observe that the spreading behaviour is not being affected by the varying particle type, size and aspect ratio. Despite using round shaped glass particles and elongated rice grains (1.0 < Aspect Ratio < 4.4), the dimensionless outer diameter of the settled shape only varied over a close-range of 0.742 ± 0.035 (4.7 % deviation; see fig. 11b). We can speculate that the macroscopic identity of the cluster was the reason why the dimensionless diameter of the settled cluster was independent of the shape, size, and types of the constituent particles. Though the temporal evolution of the cluster was affected to some extent with different sized and shaped particle used for settling, the hemispherical bottom shape of the cluster (which is the crucial criteria for the symmetric shape formation) was not affected, despite using particles of different shape, size, and type.

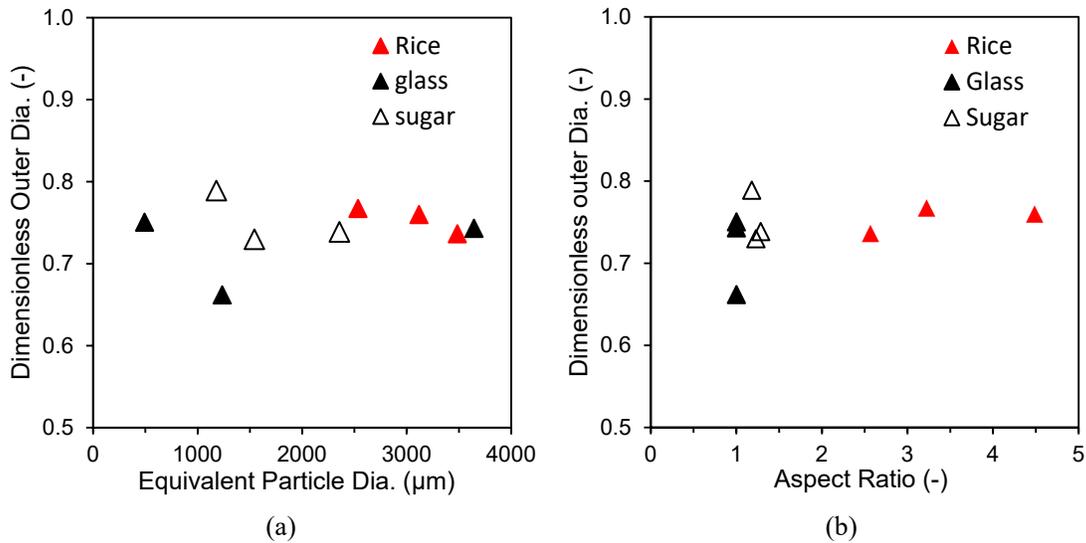

(a)                                  (b)

Fig. 11. Effect of (a) equivalent particle diameter and (b) aspect ratio of different particle types on the dimensionless outer diameter of the settled pattern

### 3.5 Effect of fluid viscosity

Cluster settling experiments were performed with ambient liquids of increasing viscosity. The settling velocity of the cluster was found to decrease significantly with increasing viscosity of the liquid, and even the spreading time was observed to increase. This is due to higher friction/ resistance provided by the liquid to the particle cluster. The $Re_c$ was also found to decrease significantly as the viscosity of the ambient liquid was increased. The dimensionless diameter of the settled shape also decreased with increased viscosity until a certain viscosity value, and then it was found to increase. The existence of a limiting viscosity (~100 cP) was witnessed beyond which no near-spherical cluster formation was observed. When the fluid viscosity is increased over the limiting viscosity value, destabilization of the cluster was found to occur, and the cluster was not able to maintain spherical shape (see fig. 12a).

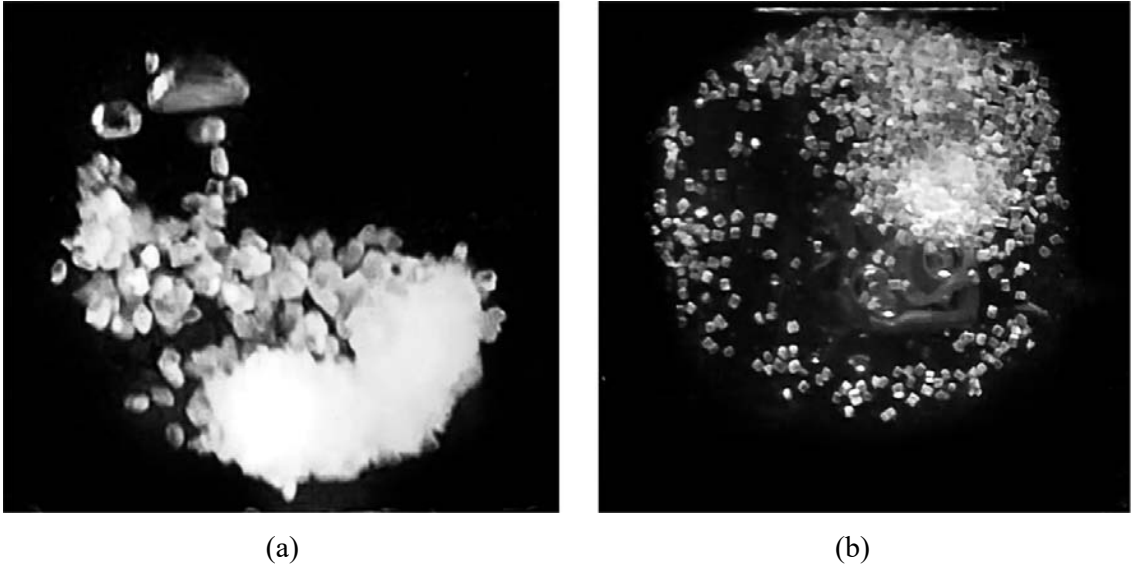

|(a)|(b)|

Fig. 12. (a) Breakage of particle cluster; (b) asymmetric spreading of the particles

As the hemispherical bottom cap of the cluster was not formed, when hitting the vessel bottom surface, the constituent particles of the cluster spread in an asymmetric manner (see fig. 12b). A significant deviation from a circular settled structure was observed. The initial decrease in the dimensionless diameter was due to a lesser density/intensity of the particulate gravity current, but after a limiting viscosity value, due to asymmetric spreading, the roundness and the dimensionless diameter of the settled shape, both increased (see fig. 14a). The limiting viscosity value (≈100 cP) of the ambient liquid was required for the fluid in order to form a cluster with a hemispherical cap, which resulted in a uniform settled structure. This is probably due to a regime change of cluster motion has occurred around the cluster Re of 60 (see fig. 14b).

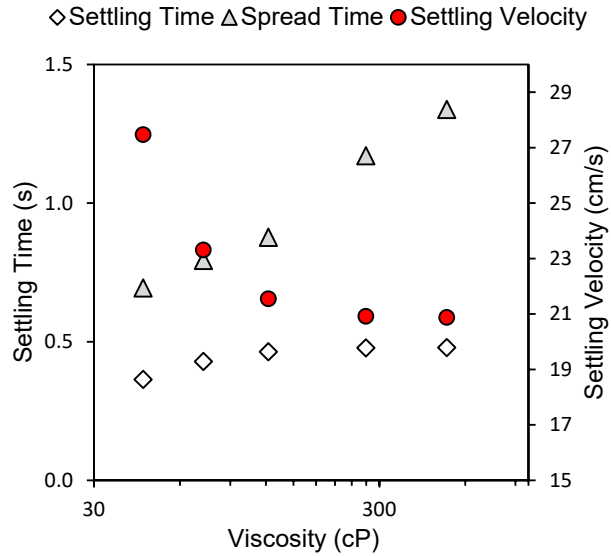

Fig. 13. Effect of fluid viscosity on the settling time, spread time and cluster settling velocity

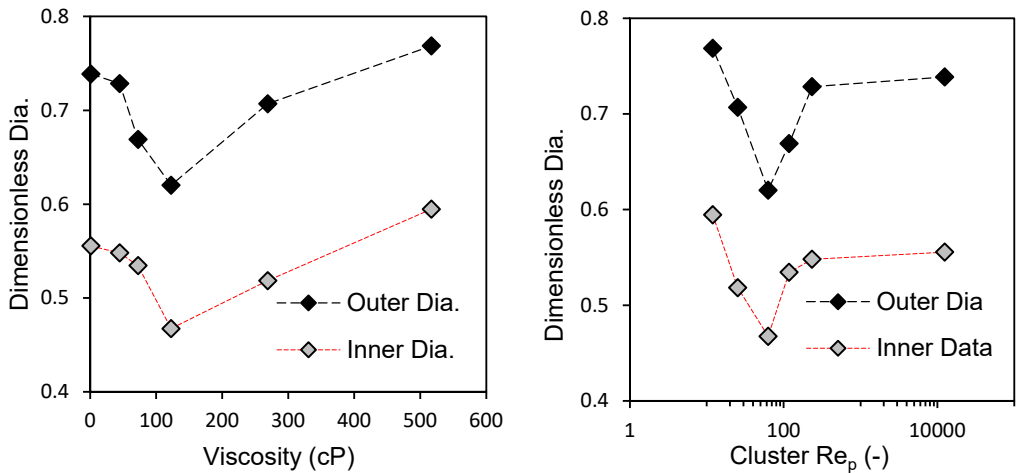

Fig. 14. (a) Effect of fluid viscosity on the dimensionless diameter of the settled pattern; (b) Effect of $Re_c$ on the dimensionless diameter of the settled pattern

## 4. Conclusion

By performing experimental investigations, we have examined the nature of settling of a cluster of particles falling in a quiescent liquid at high $Re_c$ conditions, and it's spreading at the vessel bottom surface. The significant finding is that a cluster having an initially hemispherical bottom cap settles in an axisymmetric shape upon hitting the vessel bottom surface. The formation of the hemispherical cap and the subsequent axisymmetric shape formation is found to be a robust feature of the system as it is found to nearly independent of the particle shape, size, and types. With the increasing initial mass of the particles, the cluster size was found to increase, the tendency for tail formation was reduced, and the settled shape became closer to a circle and of higher dimensionless diameter. Despite using vessels of different sizes, the settled shape was close to a circle. However, the square shape vessels resulted in enhanced spreading behavior. It is worth noting that a limiting viscosity of the ambient liquid of 100 cP was observed beyond which the cluster shape was deformed, and the particles spread in an asymmetric manner. The experimental analysis presented here should be taken as a step towards understanding the effect of high Reynolds number hydrodynamic interactions between settling particle clusters on its spreading behaviour upon hitting the bottom surface.

## Acknowledgment

Sayan Pal acknowledges the Council of Scientific and Industrial Research (CSIR) for PhD. fellowship. The authors thank the Dept. of Science & Tech. (GoI) for the funding through Swarnajayanti Fellowship DST/SJF/ETA-03/2014-15. The authors thank Mr. Jaysingh Yadav for helping in performing the experiments.